# Tuning Excited State Electron Transfer in Fe Tetracyano-Polypyridyl Complexes

Kristjan Kunnus,[1] Lin Li,[1] Charles J. Titus,[2] Sang Jun Lee,[3] Marco E. Reinhard,[1] Sergey Koroidov,[1] Kasper S. Kjær,[1] Kiryong Hong,[1] Kathryn Ledbetter,[1] William B. Doriese,[4] Galen C. O'Neil,[4] Daniel S. Swetz,[4] Joel N. Ullom,[4] Dale Li,[3] Kent Irwin,[2,3] Dennis Nordlund,[3] Amy A. Cordones,[1] Kelly J. Gaffney[1]

**Affiliations**

[1] Stanford PULSE Institute, SLAC National Accelerator Laboratory, Stanford University, Menlo Park, California 94025, United States.

[2] Department of Physics, Stanford University, Stanford, California 94305, USA.

[3] SLAC National Accelerator Laboratory, Menlo Park, California 94025, United States.

[4] National Institute of Standards and Technology, Boulder, CO 80305, USA

*Abstract*

We have investigated photoinduced intramolecular electron transfer dynamics following metal-to-ligand charge-transfer (MLCT) excitation of $[Fe(CN)_4(2,2'\text{-bipyridine})]^{2-}$ (**1**), $[Fe(CN)_4(2,3\text{-bis(2-pyridyl)pyrazine})]^{2-}$ (**2**) and $[Fe(CN)_4(2,2'\text{-bipyrimidine})]^{2-}$ (**3**) complexes in various solvents with static and time-resolved UV-visible absorption spectroscopy and Fe 2p3d resonant inelastic X-ray scattering. We observe $^3$MLCT lifetimes from 180 fs to 67 ps over a wide range of MLCT energies in different solvents by utilizing the strong solvatochromism of the complexes. Intramolecular electron transfer lifetimes governing $^3$MLCT relaxation increase monotonically and (super)exponentially as the $^3$MLCT energy is decreased in **1** and **2** by changing the solvent. This behavior can be described with non-adiabatic classical Marcus electron transfer dynamics along the indirect $^3$MLCT→$^3$MC pathway, where the $^3$MC is the lowest energy metal-centered (MC) excited state. In contrast, the $^3$MLCT lifetime in **3** changes non-monotonically and exhibits a maximum. This qualitatively different behaviour results from direct electron transfer from the $^3$MLCT to the electronic ground state (GS). This pathway involves nuclear tunnelling for the high-frequency polypyridyl skeleton mode ($\hbar\omega \approx 1530$ cm$^{-1}$), which is more displaced for **3** than for either **1** or **2**, therefore making the direct pathway significantly more efficient in **3**. To our knowledge, this is the first observation of an efficient $^3$MLCT→GS relaxation pathway in an Fe polypyridyl complex. Our study suggests that further extending the MLCT excited state lifetime requires (1) lowering the $^3$MLCT state energy with respect to the $^3$MC state and (2) suppressing the intramolecular distortion of the electron-accepting ligand in the $^3$MLCT excited state to suppress the rate of direct $^3$MLCT→GS electron transfer.

*Introduction*

Transition metal complexes with strong charge transfer absorption bands in the visible spectral region can be utilized as photosensitizers for solar energy conversion applications.[1] Conventional molecular photosensitizers use noble metals, such as Ru, leading to higher costs and motivating the identification of photosensitizers using abundant metals, such as Fe.[2–5] Development of Fe-based photosensitizers has been slow due to the rapid deactivation of the light absorbing metal-to-ligand charge-transfer (MLCT) states, which occurs with unity quantum yield within ~100 fs for the Fe analogs of the conventional Ru polypyridyl photosensitizers, and results in the loss of electronic energy into heat.[6–8] Extending the lifetime of MLCT excited states has, therefore, been recognized as one of the key challenges in developing functional Fe-based photosensitizers. We address this goal with an experimental investigation of the MLCT relaxation pathways of three solvatochromic tetracyano-polypyridyl photosensitizer model complexes **1**, **2**, and **3** (Scheme 1) in a large range of solvents. The goals of this work are to systematically investigate the dependence of MLCT lifetime on the MLCT energy and to understand the forces governing MLCT deactivation dynamics within the framework of established electron transfer models.

A range of synthetic strategies have been put forward to achieve Fe complexes with extended MLCT lifetimes.[9] McCusker *et al.* destabilized the metal-centered (MC) states of tridentate terpyridine complex by introducing a bridging carboxyl ligand between pyridyl units, leading to higher symmetry and π-back-bonding.[10,11] The ligand field increase was not sufficient to significantly increase the MLCT lifetime and the complex decays to a $^5$MC state on an ultrafast timescale.[12] Damrauer *et al.* pursued a different approach by introducing halogen substituents to terpyridine ligands to decrease the ligand field and sterically hinder the motions facilitating MLCT→MC relaxation.[13,14] These complexes have high spin ground states and $^{5,7}$MLCT excited states with lifetimes up to 17.4 ps. The most successful approach to date was introduced by Wärnmark *et al.*[15,16] and followed by Gros *et al.*[17] This utilizes strongly σ-donating N-heterocyclic carbene ligands to destabilize the MC excited states. This strategy yielded 92% electron injection efficiency to $TiO_2$ from an Fe complex with tridentate pyridine-carbene ligands and with an intrinsic 9 ps MLCT lifetime.[18] Chabera *et al.* reported an all-carbene six-coordinated ferric complex with a ligand-to-metal charge-transfer (LMCT) lifetime of 100 ps[19] and an analogous ferrous complex with the longest MLCT lifetime detected so far: 528 ps.[20] This work has been extended by Kjær *et al.*, who have demonstrated a 2 ns lifetime LMCT state in a near-octahedral Fe(III) complex with two tri-dentate carbene-borate ligands that can reduce the methylviologen cation and oxidize diphenylamine.[21]

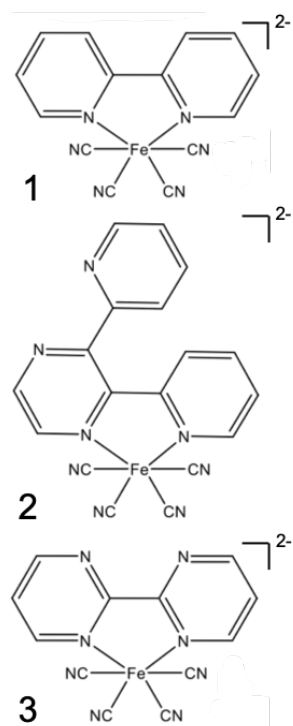

Scheme 1. Structural formulas of the complexes **1**, **2** and **3**.

Our approach is inspired by early visible transient absorption experiments on complex **1** by Winkler *et al.*[22] and by the numerous investigations on heteroleptic Ru cyano-polypyridyl complexes.[23–25] The introduction of strongly coordinating cyanide ligands serves the purpose of increasing the ligand field, thus destabilizing the MC excited states and slowing the MLCT→MC relaxation pathways. To this end, photosensitization of $TiO_2$ with Fe cyano-bipyridine complexes has been demonstrated, although with a very low yield.[26,27] We have previously shown with femtosecond X-ray emission spectroscopy (XES) that while [Fe(CN)$_2$(bpy)$_2$] MLCT excited states relax to $^5$MC states within ~100 fs,[28] similar to [Fe(bpy)$_3$]$^{2+}$ (bpy=2,2'-bipyridine),[8,29–32] complex **1** in dimethylsulfoxide (DMSO) has a MLCT lifetime of ~19 ps.[33] The MLCT energies of **1** are strongly solvatochromic[34] and this also has a significant effect on the MLCT lifetime. In a recent study using UV-visible transient absorption and femtosecond XES, we established that the MLCT lifetime of **1** in $H_2O$ is shortened to ~100 fs and the observed 13 ps lifetime intermediate corresponds to a $^3$MC state.[35] Motivated by these results, the current investigation aims to systematically track the MLCT energy dependence of the photoinduced dynamics in tetracyano-pyridyl Fe complexes. In particular, we explore the impact that further lowering of the MLCT energy has on the MLCT lifetime. For the latter reason, we have widened our study to include complexes **2** and **3** that show red-shifted MLCT absorption bands relative to **1** (Fig. 1).

MLCT lifetimes in Fe coordination complexes are determined by intramolecular electron transfer. Fundamental concepts in any physical theory of electron transfer are driving force $\Delta G$ and reorganization energy $\lambda$:

$$\Delta G \equiv E_0(P) - E_0(R),$$
$$\lambda \equiv E_R(P) - E_0(P),$$

here $E_0(R)$, $E_0(P)$ are the energies of the reactant and product states at their optimal geometries and $E_R(P)$ is the energy of the product state at the optimal reactant geometry. If both reactant and product surfaces are harmonic with identical force constants, and the electronic coupling is weak (non-adiabatic approximation), then these two parameters determine the minimum energy barrier

$$\Delta G^{\#} = \frac{(\Delta G + \lambda)^2}{4\lambda}.$$

The energy barrier $\Delta G^{\#}$, together with the temperature T and the pre-exponential factor $k_0$, uniquely define the electron transfer rate $k$ in the case of classical (quasi)equilibrium nuclear dynamics

$$k = k_0 exp\left(-\frac{\Delta G^{\#}}{k_B T}\right).$$

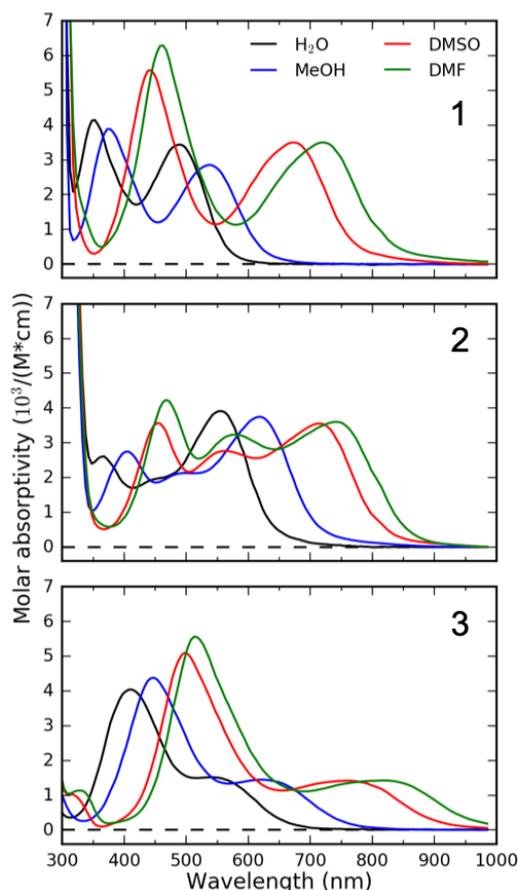

Fig. 1. UV/vis absorption spectra of **1**, **2** and **3** in selected solvents demonstrating solvatochromism of metal-to-ligand charge-transfer (MLCT) bands. MeOH – methanol, DMSO – dimethylsulfoxide, DMF – dimethylformamide.

These considerations form the basis of the Marcus electron transfer theory.[36,37] In the case of intramolecular electron transfer, it is generally necessary to consider nuclear motions quantum mechanically and this has given rise to treatments based on Fermi's Golden Rule, pioneered by Bixon and Jortner.[38,39] The non-adiabatic electron transfer rate is then given as

$$k = \frac{2\pi |V|^2}{\hbar} \sum_M \sum_{i,j} p_M(i) FC_M(i,j)\, \delta(E_i - E_j),$$

where $V$ is electronic coupling between the reactant and product states, $p_M(i)$ is Boltzmann distribution for mode M in the reactant state, $FC_M(i,j)$ is a Franck-Condon factor between reactant and product vibrational states i and j for mode M. Specifically, electron transfer in metal complexes between the metal and ligand typically activates high-frequency ligand vibrations ($h\nu > k_B T$). Quantum effects can be particularly important in the inverted Marcus region ($-\Delta G > \lambda$),[40,41] as in the case of direct $^3$MLCT→GS electron back-transfer of Ru and Os polypyridyl complexes.[42,43] The nano- and microsecond $^3$MLCT lifetimes in these metal photosensitizers can be often described by the energy gap law.[44] This contrasts to femto- and picosecond $^3$MLCT lifetimes of Fe polypyridyl complexes where the electron back-transfer follows an indirect $^3$MLCT→MC→GS pathway. As opposed to the thoroughly investigated direct pathway of Ru/Os complexes, the indirect pathway in Fe complexes is considerably less understood. Here we investigate the forces governing the indirect electron transfer pathway by systematically modifying the MLCT energies through solvatochromism. With this we aim to determine the relative energetics of the MLCT and MC potential energy surfaces ($\Delta G$ and $\lambda$) and connect these to the observed electron transfer rates. Establishing the latter relationship is not only necessary for understanding the forces governing electron back-transfer, but it is also desirable for further rational design of MLCT lifetimes and photoredox properties. In addition to the indirect pathway, we will also evaluate the rates of the direct $^3$MLCT→GS pathway. To the best of our knowledge, we report here a first Fe complex where the $^3$MLCT relaxation is dominated by the direct pathway.

Many energetic parameters relevant for electron transfer processes can be extracted from UV-Visible absorption spectra. This approach has been successfully employed in the investigations of inter-valence electron transfer rates in mixed-valence complexes, where the relevant parameters can be retrieved from Mulliken-Hush analysis.[45–48] The situation is more complicated in the present case, because the MLCT→MC electron transfer takes place between the excited states of the molecule. UV-Visible absorption spectra of Fe polypyridyl complexes report only about the energetics of reactant MLCT states and do not contain direct information about the relevant MC product states. Dipole transitions to the relevant MC states are Laporte and spin-forbidden, and they overlap energetically with the dipole allowed MLCT absorption bands. In order to access the energetics of the relevant MC excited states we have utilized resonant inelastic X-ray scattering (RIXS) at the Fe $L_3$-edge (700–715 eV).

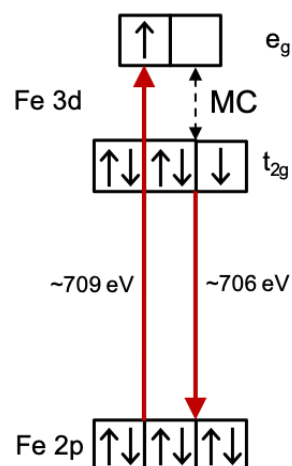

Scheme 2. Resonant inelastic X-ray scattering (RIXS) at the Fe L-edge $e_g$-resonance of an octahedral low-spin Fe $3d^6$ complex. Dominant RIXS final states are metal-centered (MC) excited states.

RIXS provides a powerful spectroscopic technique for measuring low-energy excitations of matter with the element selectivity of X-ray spectroscopy.[49] RIXS at 3d transition metal L-edges involves resonant dipole allowed 2p→3d excitation and de-excitation transitions (Scheme 2). The RIXS final states do not contain a core-hole and the resulting RIXS spectra are dominated by relevant MC valence excited states.[50] Additionally, because MC and MLCT states have electrons excited to different unoccupied molecular orbitals, different final excited states can be accessed by tuning the incident X-ray photon energy to different X-ray absorption resonances. Therefore, a RIXS spectrum measured at the metal L-edge 3d absorption resonance (white line) is dominated by the MC final states, while being completely free of MLCT final states.[51] This makes metal L-edge RIXS a highly suitable probe of MC excited state energetics, a capability we utilize in the present work.

The application of RIXS to chemically relevant systems, often with dilute concentrations and of limited sample quantity, is currently significantly constrained due to the small cross-sections of RIXS processes and the low efficiencies of soft X-ray grating spectrometers. Below we will demonstrate RIXS spectra measured with novel X-ray spectrometers based on transition-edge sensor (TES) technology[52–55] These energy dispersive spectrometers have >100 times higher detection efficiency compared to conventional grating spectrometers. TES spectrometer employed in the current work has 2.3 eV (FWHM) energy resolution, We demonstrate that this is sufficient to spectrally separate elastic scattering and inelastic MC and charge transfer states scattering channels in the complexes studied here. Although the current energy resolution of TES spectrometers still limits its applicability to valence electronic structure investigations, developments presently underway aim to improve their resolution below 1 eV.

*Experimental*

**Samples.** We purchased the potassium salts of complexes **1**, **2** and **3** from Allichem Inc. and used without further purification.[56] Data from the same synthesis of complex **1** have been published before.[28,33] Complexes **2** and **3** were examined with elemental analysis: **2** (C18H16O3FeK2N8) Calc: C, 37.25; H, 3.82; N, 19.30%. Found: C, 37.04%; H, 2.34%; N, 19.50%. **3** (C12H12O3FeK2N8) Calc: C, 32.01; H, 2.69; N, 24.88%. Found: C, 31.81%; H, 2.18%; N, 24.63%. We exchanged the $K^+$ with tetrabutylammonium ($TBA^+$) to increase the solubility in organic solvents. Firstly, the $H^+$-form of complexes were prepared following a published procedure.[57] The resulting product was reacted with a TBA-OH (Sigma-Aldrich) methanol solution in a stoichiometric amount, which was subsequently dried to retrieve the complexes in the $TBA^+$-form.[16]

**UV/Vis absorption.** We conducted femtosecond time-resolved UV-Visible transient absorption (TA) and steady-state UV-Visible absorption measurements on various solutions of complexes **1**, **2** and **3**. We used $K^+$-salts for aqueous solutions and $TBA^+$-salts for all other solutions. Complex **1** was studied in $H_2O$, methanol (MeOH), dimethysulfoxide (DMSO), acetonitrile (MeCN) and dimethylformamide (DMF). Complex **2** was studied in $H_2O$, MeOH, DMSO and DMF. Complex **3** was studied in $H_2O$, MeOH, butanol, hexanol, dichloromethane (DCM), propylene carbonate, pyridine, benzonitrile, DMSO, MeCN, acetone and DMF. All solvents were reagent grade. Concentrations of all the solutions were adjusted to have the maximum absorption in the visible region between 0.3–0.5 OD for a 100 μm path length (concentrations of a few mM). For TA measurements of aqueous solutions we used a recirculating sheet jet with 100 μm thickness. For all the other samples we used a vibrating 100 μm thick quartz cell without flowing the solution. We measured all steady state UV-Visible absorption spectra using a 100 μm quartz cell with a Cary 50 spectrophotometer.

TA experiments were carried out using an amplified Ti:sapphire laser system (Coherent Mantis or Vitara oscillator with Coherent Legend Elite Duo) with a 5 kHz repetition rate, 800 nm central wavelength, 2 mJ pulse energy and 40 fs FWHM pulse duration. A portion of the laser pumped an optical parametric amplifier (Spectra-Physics OPA-800C) to generate a near IR signal and idler. We used sum-frequency generation of the signal and 800 nm light to make 500–550 nm pump pulses and frequency doubling of the signal to generate 610–820 nm pump pulses. The pump pulse was directed to the sample through a delay stage, a 2.5 kHz chopper, and a lens, resulting in pump pulses with 200–300 μm focus diameter (FWHM) and fluences of 1–5 mJ/cm$^2$. The pump was overlapped with a white light probe pulse (via supercontinuum generation in 4 mm of $CaF_2$) at the sample position. The probe was transmitted through the sample and imaged on a spectrometer (Horiba Jobin Yvon iHR320, grating 150 grooves/mm). The probe spectrum was recorded at 5 kHz with a NMOS linear image sensor (Hamamatsu, S8380-512Q) simultaneously over the whole 350–750 nm spectral range. The instrument response function was estimated to be ~100 fs (FWHM). We measured UV-Visible absorption spectra of the solutions before and after the TA experiments to check for photodamage. Most samples showed no optical degradation, with a few exhibiting a few percent reduction in absorption. The differential absorbance (ΔA) was calculated as ΔA=log($I_{off}/I_{on}$), where $I_{on}$ and $I_{off}$ are the pumped and unpumped intensity, respectively. We used the cross-phase modulation signal between the pump and the probe to determine $t_0$ for each probe wavelength and to correct for the group velocity dispersion.

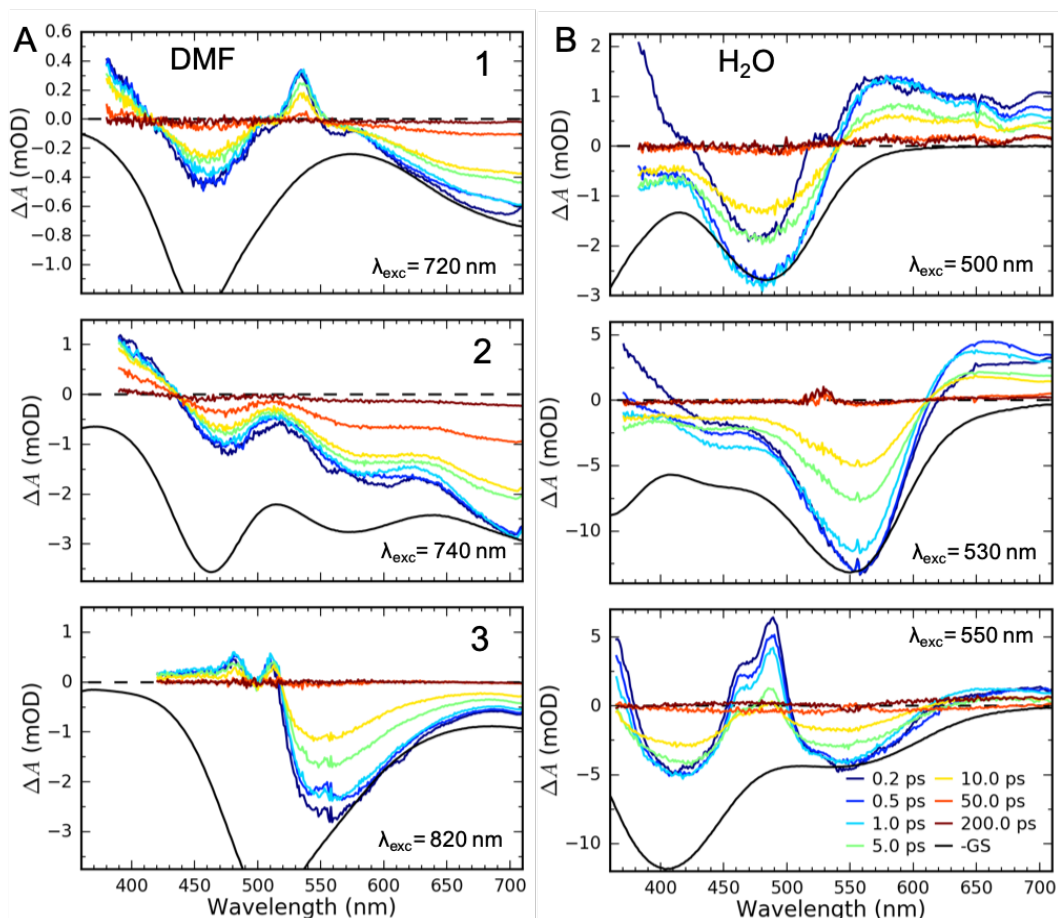

Fig. 2. Femtosecond transient absorption (TA) data of **1**, **2** and **3** in (A) DMF and (B) H$_2$O. Negative ground state (GS) spectra (in black) are scaled to match the TA signal amplitudes for comparison. The artefact around 530 nm for **2** in water results from pump scatter.

**Resonant Inelastic X-ray Scattering.** RIXS experiments were carried out at beamline 10-1 at the SSRL storage ring at SLAC National Accelerator Laboratory. The endstation and soft X-ray TES spectrometer[55] are described in more detail by Titus *et al.*[58] The employed TES signal processing techniques are described by Fowler *et al.*[59] Powder samples of K$^+$-salt complexes were pressed on a carbon tape attached to a sample holder. Samples were kept at room temperature. RIXS maps over the complete Fe L$_{2,3}$-edge were collected by scanning the incident x-ray energy from 700–735 eV (0.1 eV step, 0.2 eV monochromator bandwidth) and raster scanning over 20 spots on each sample (two monochromator scans on each spot, 1 mm × 1 mm X-ray footprint, average incident flux ~5·10$^{10}$ photons/s). No beam damage effects were observed in the collected data. The RIXS maps of each sample were acquired for 2 h and 45 min. The incident photon energy was calibrated to match the Fe L$_3$-edge e$_g$-resonance of K$_4$[Fe(CN)$_6$] published by Hocking *et al.*[60]

X-ray photon energies detected by the TES spectrometer were calibrated with a procedure described in Ref. 61 that gives 0.4 eV uncertainty in absolute photon energies. This calibration was refined for Fe 2p3d emission by shifting the elastically scattered photon energy to match the incident photon energy. We estimate the uncertainty in relative photon energies within is below 0.1 eV. The spectral response of the TES spectrometer was measured at 750 eV with elastically scattered light from a gold film (see Supporting Information, SI).

*Results*

Below we present the results from three different sets of experiments. Firstly, we performed femtosecond pump-probe UV-Visible transient absorption measurements to determine the excited state relaxation dynamics and the MLCT lifetimes. Secondly, we carried out band shape analysis of the steady state UV-Visible absorption spectra to quantify the MLCT excitation energies and the associated intra- and intermolecular reorganization energies. Thirdly, we performed steady state Fe 2p3d RIXS experiments to establish the MC excited state energies. Results from each of these experiments are described in detail in the following sub-sections.

**UV-Visible transient absorption.** UV-Visible transient absorption (TA) experiments were used to probe the photoexcited MLCT state relaxation dynamics of **1**, **2** and **3** in various solvents with different solvatochromic effects. In Fig. 2, we display time-resolved difference TA spectra of **1**, **2** and **3** in DMF and $H_2O$. Of all the studied solvents, these have the largest relative solvatochromic effect, with the DMF and $H_2O$ solutions being most red- and blue-shifted, respectively. The complete collection of measured TA data is included in the Supporting Information (SI).

Characteristic absorption signatures of the MLCT excited states are well established for transition metal polypyridyl complexes, and these are very similar to the absorption features of reduced polypyridyl radicals.[62,63] This facilitates robust assignment of MLCT relaxation dynamics. We assign the excited state absorption (ESA) feature positioned below 550 nm for all of the measured datasets to intra-ligand transitions of the reduced polypyridyl radicals in the lowest energy MLCT states of the complexes.[64] These MLCT ESA features decay concomitantly with the ground state bleach (GSB) recovery in most of the studied solutions of all three complexes (Fig. 2, except **1** and **2** in $H_2O$). This concomitant decay of ESA and GSB features confirms only the MLCT excited state is significantly populated. This is consistent with the previously published results of **1** in acetone and DMSO.[22,33]

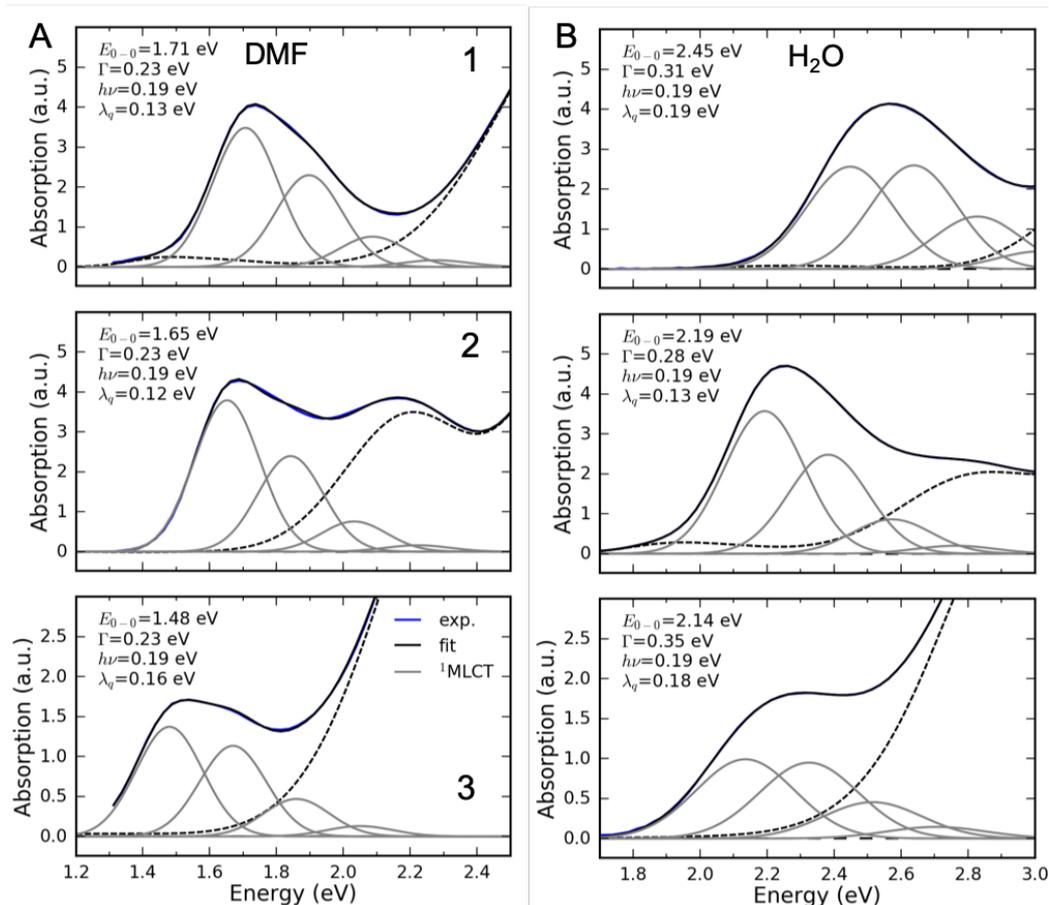

Fig. 3. Fits of the lowest $^1$MLCT UV-Visible absorption peak of **1**, **2** and **3** in (A) DMF and (B) $H_2O$. Energetic parameters of the lowest $^1$MLCT peak retrieved from a fit are displayed in each panel. $E_{0-0}$ – energy of the zero-phonon peak, $\Gamma$ – Gaussian broadening (FWHM), $\lambda_q$ – reorganization energy of the high frequency mode, and $h\nu$ – quantum energy of the high frequency mode. Black dashed lines show residual absorption that is not due to the lowest energy $^1$MLCT state.

Qualitatively different dynamics are observed in $H_2O$ and MeOH solutions of **1** and **2** where the MLCT ESA features associated with the polypyridyl radical decay on the sub-picosecond timescale without GSB recovery. Our recent work combining TA with time-resolved Fe Kα/Kβ X-ray emission spectroscopy (XES) measurements of **1** in $H_2O$ determined that the MLCT state decays into a triplet metal-centred ($^3$MC) excited state.[35] This allows us to unambiguously assign also the ~10 ps intermediate of **2** in $H_2O$ and MeOH solutions to a $^3$MC excited state. The intermediate lacks the signature ESA of a reduced ligand, while exhibiting the broad ESA feature to the red of the GSB. We previously assigned such TA signature to the $^3$MC state.[35] We observe no population of MC states in any solution of **3**.

To accurately determine the time scales of the observed population dynamics, we carried out a global fitting of the TA data. The singular value decomposition-based kinetics fitting procedure utilized here has become a standard method for the analysis of 2D TA data.[65] A description of the procedure and all of the fit results are presented in the SI. A summary of the extracted MLCT lifetimes is presented in Table 1. **1** and **2** show more than two orders of magnitude decrease in the MLCT lifetime as the MLCT energy is increased by more strongly interacting solvents. MLCT lifetimes decrease from 29 ps to 200 fs for complex **1**, and from 67 ps to 300 fs for **2**. The 67 ps MLCT lifetime of **2** in DMF is the longest MLCT lifetime observed in polypridyl-containing Fe complexes.

Table 1. Summary of the vibronic band shape analysis of the lowest energy $^1$MLCT UV-Visible absorption band and the MLCT excited state lifetimes extracted from transient absorption (TA) experiments. $E_v$ and $E_0$ correspond to the lowest $^1$MLCT energy at the GS and MLCT geometries, respectively. $\lambda_{cl}$ and $\lambda_q$ are GS-$^1$MLCT reorganization energies for the classical and high-frequency quantum modes, respectively. $\tau_{MLCT}$ is the MLCT lifetime retrieved from the global fitting of TA data.

| Complex | Solvent | $E_v$ (eV) | $\lambda_{cl}$ (eV) | $\lambda_q$ (eV) | $E_0$ (eV) | Excitation energy nm | Excitation energy eV | $\tau_{MLCT}$ (ps) |
|---|---|---|---|---|---|---|---|---|
| **1** | $H_2O$ | 2.64 | 0.33 | 0.19 | 2.12 | 500 | 2.48 | 0.18 |
| | MeOH | 2.41 | 0.32 | 0.17 | 1.92 | 520 | 2.38 | 0.22 |
| | DMSO | 1.95 | 0.20 | 0.13 | 1.62 | 670 | 1.85 | 16.5 |
| | MeCN | 1.94 | 0.20 | 0.13 | 1.61 | 720 | 1.73 | 19.3 |
| | DMF | 1.83 | 0.19 | 0.12 | 1.52 | 720 | 1.73 | 28.7 |
| **2** | $H_2O$ | 2.33 | 0.28 | 0.13 | 1.92 | 530 | 2.34 | 0.30 |
| | MeOH | 2.10 | 0.24 | 0.12 | 1.73 | 610 | 2.03 | 0.61 |
| | DMSO | 1.83 | 0.19 | 0.11 | 1.53 | 720 | 1.73 | 31.6 |
| | DMF | 1.77 | 0.19 | 0.12 | 1.47 | 740 | 1.68 | 66.9 |
| **3** | $H_2O$ | 2.32 | 0.42 | 0.18 | 1.71 | 550 | 2.25 | 11.9 |
| | MeOH | 2.04 | 0.33 | 0.16 | 1.55 | 625 | 1.98 | 16.9 |
| | Butanol | 1.88 | 0.25 | 0.16 | 1.48 | 690 | 1.80 | 20.0 |
| | Hexanol | 1.76 | 0.24 | 0.16 | 1.36 | 670 | 1.85 | 19.9 |
| | DCM | 1.79 | 0.23 | 0.16 | 1.40 | 730 | 1.70 | 20.7 |
| | Propylene carbonate | 1.74 | 0.23 | 0.16 | 1.36 | 750 | 1.65 | 21.4 |
| | Pyridine | 1.76 | 0.24 | 0.16 | 1.36 | 750 | 1.65 | 22.6 |
| | Benzonitrile | 1.75 | 0.24 | 0.16 | 1.35 | 760 | 1.63 | 18.5 |
| | DMSO | 1.73 | 0.22 | 0.15 | 1.36 | 775 | 1.60 | 19.4 |
| | MeCN | 1.73 | 0.24 | 0.15 | 1.34 | 775 | 1.60 | 13.5 |
| | Acetone | 1.70 | 0.25 | 0.16 | 1.29 | 775 | 1.60 | 16.6 |
| | DMF | 1.64 | 0.19 | 0.16 | 1.29 | 820 | 1.51 | 12.7 |

Surprisingly, complex **3** exhibits qualitatively different behaviour. The MLCT lifetime dependence of **3** on the MLCT energy is non-monotonic and changes only by a factor of two, staying within a range of 12–23 ps in all solvents (Table 1). The MLCT lifetime increases from ~12 ps at the lowest MLCT energy in DMF and reaches a maximum in pyridine solution (~23 ps), and then it decreases again to ~12 ps in $H_2O$. We provide an explanation to this qualitatively different behaviour of **3** in the Discussion section.

**UV-Visible absorption.** We carried out a vibronic band shape analysis of the lowest-energy MLCT UV-visible absorption features to obtain the MLCT energies and the reorganization energies associated with the GS→MLCT transitions in all the investigated solutions. Similar vibronic band shape analysis has been successfully applied to the MMCT bands of mixed-valence complexes[66] and MLCT bands of Fe and Ru polypyridyl complexes.[67–69] We model the lowest energy $^1$MLCT absorption with a single high-frequency quantum mode and with an effective classical mode that includes all the low-frequency modes. The vibronic shape of the MLCT absorption band is then described by

$$I(E) = \sum_n \frac{S^n e^{-S}}{n!} exp\left[-4ln(2)\frac{(E - E_{0-0} - nh\nu)^2}{\Gamma^2}\right],$$

where $S=\lambda_q/h\nu$ is the Huang-Rhys factor for the high-frequency quantum mode, and $\lambda_q$ and $h\nu$ are the respective reorganization energy and vibrational quanta energy. $E_{0-0}$ is the zero-phonon energy of high-frequency vibration. The gaussian broadening $\Gamma$ (FWHM) is related directly to the classical reorganization energy:

$$\lambda_{cl} = \frac{\Gamma^2}{16 \ln 2 \, k_B T}.$$

The fully relaxed (minimum) energy of the $^1$MLCT state $E_0$ with respect to the ground state is $E_0 = E_{0-0} - \lambda_{cl} = E_v - \lambda_q - \lambda_{cl}$, where $E_0$ and $E_v$ correspond to the optimal and vertical $^1$MLCT energies, respectively. Thus, the energetic parameters that are directly fitted to the lowest energy $^1$MLCT peak are $E_{0-0}$, $\Gamma$, $\lambda_q$ and $h\nu$ (Fig. 3, Table 1). These fit parameters provide the required input to calculate $\lambda_{cl}$, $E_0$, and $E_v$, all of which are reported in Table 1.

The application of the vibronic band shape formula is experimentally justified by close inspection of UV-Visible absorption spectra in DMF (Fig. 3). There the $^1$MLCT peak has a pronounced high energy shoulder (this is clearly observed also in other solutions, e.g. DMSO and MeCN; see the SI), and fitting yields $h\nu$ = 0.19±0.01 eV (1530±80 $cm^{-1}$). Within the experimental uncertainty, we find that $h\nu$ is the same in all studied solutions within experimental error. Therefore, in the solutions where vibronic structure is not resolved, due to larger Gaussian broadening and/or overlapping higher energy MLCT peaks, we have fixed $h\nu$ = 0.19 eV. Note that this vibrational frequency agrees well with the polypyridyl skeleton mode observed in the previous studies of Ru and Fe polypyridyl complexes.[67,70,71] For example, a 0.199 eV (1607 $cm^{-1}$) high frequency mode dominates the vibronic structure of lowest energy $^1$MLCT band of $[Fe(bpy)_3]^{2+}$.[67]

In several measured UV-Visible absorption spectra a low-energy tail is present that cannot be described by the $^1$MLCT vibronic structure or Gaussian broadening (see **1** in DMF, Fig. 3). Such low-energy features have been observed in Fe polypyridyl complexes and were assigned to the $^3$MLCT state.[67,70] In particular, Kober *et al.* carried out a detailed analysis of the low-temperature [Fe(bpy)$_3$]$^{2+}$ UV-Visible absorption spectrum and found that the $^3$MLCT state is 0.25 eV below the $^1$MLCT state with a similar vibronic structure.[70] Therefore, in the fitting of all our room temperature spectra, we have included a peak that is 0.25 eV below the $^1$MLCT and has an identical shape to the $^1$MLCT. This successfully describes the low-energy tail in all the measured spectra. The intensity of the $^3$MLCT peak is typically >50 times smaller than the $^1$MLCT intensity.

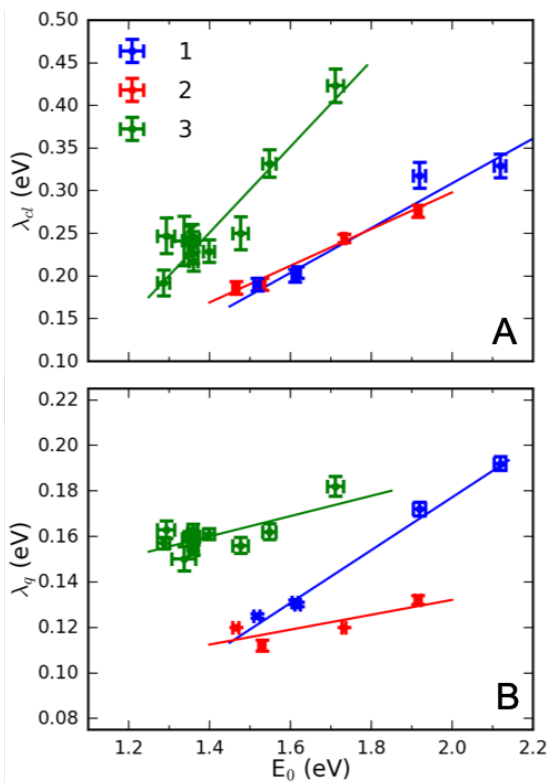

Fig. 4. Dependence of (A) classical reorganization energy and (B) reorganization energy of the high-frequency quantum mode of **1**, **2** and **3** in various solvents from the fits of the lowest energy $^1$MLCT UV-visible absorption peak.

The extracted $E_v$, $E_0$, $\lambda_{cl}$, and $\lambda_q$ values for all solutions are presented in Table 1. Several relevant trends can be observed (Fig. 4). Firstly, the solvatochromic effect is strongest in **1**, with $\Delta E_0 \equiv E_0(H_2O) - E_0(DMF) = 0.6$ eV. In comparison, $\Delta E_0$= 0.45 eV and 0.42 eV in **2** and **3**, respectively. These observations are consistent with the study of Toma *et al.*[34]. Secondly, the classical reorganization energy $\lambda_{cl}$ increases linearly with the solvatochromic effect (Fig. 4A), but the proportionality between energies is roughly a factor of two larger for **3**. Thirdly, the reorganization energy of the high frequency mode $\lambda_q$ is significant, accounting for 30–40% of the total reorganization energy. The respective Huang-Rhys factors are between 0.6 and 1. Most importantly, we observe that the reorganization energies of **3** are consistently larger than in **1** and **2**. This is particulalry evident for $\lambda_q$ in weakly interacting solvents. Although $\lambda_{cl}$ of **3** is similar to **1** and **2** in weakly interacting solvents, in H$_2$O it is significantly larger than in **1** and **2**. In the Discussion section we will show that these findings are relevant to explain the qualitatively different MLCT lifetime behavior of complex **3** from **1** and **2**.

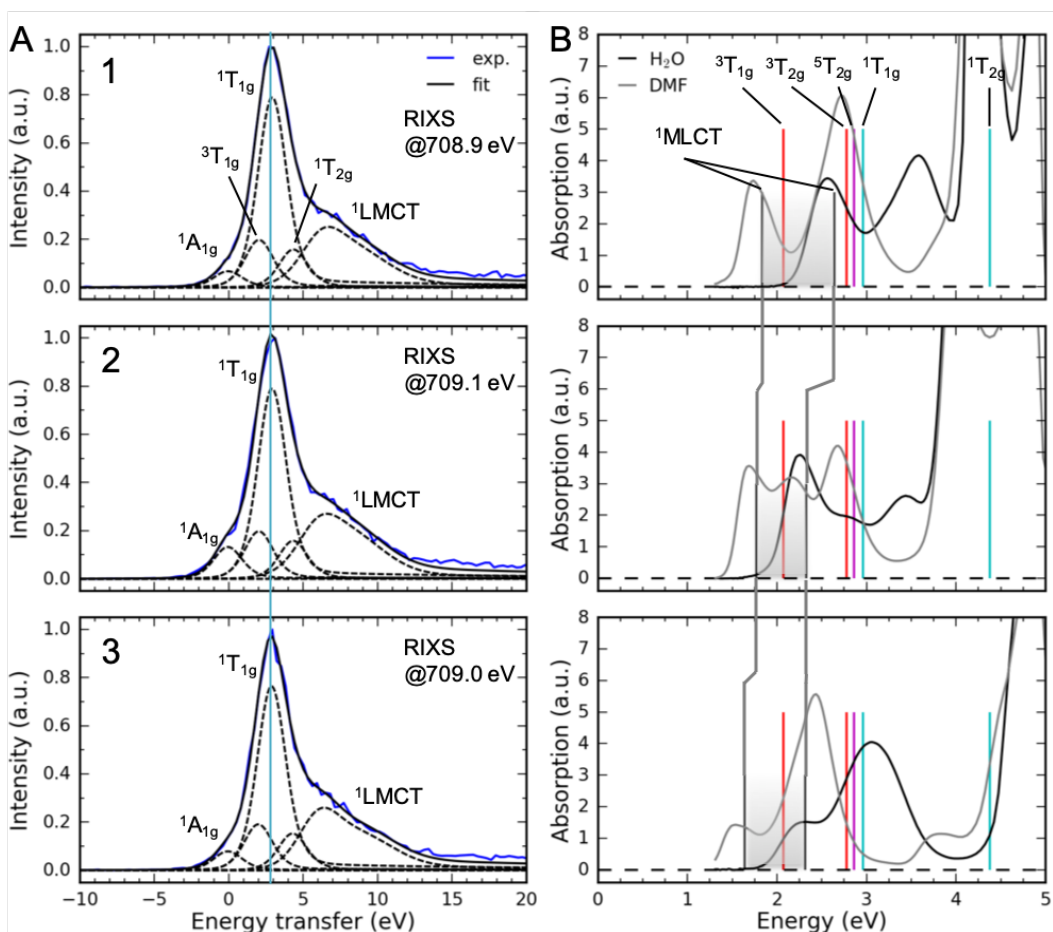

Fig. 5. (A) Fe 2p3d resonant inelastic X-ray scattering spectra at the Fe $L_3$-edge $e_g$-resonance. The light blue line corresponds to the vertical energy of the dominant $^1T_{1g}$ RIXS feature. (B) UV/vis absorption spectra and the vertical energies of MC excited states derived from RIXS (Table 2, average values). Also shown is the $^5T_{2g}$ energy at 2.86 eV. Vertical gray lines correspond to $^1$MLCT vertical energies ($E_v$, Table 1). Shaded area shows the range of vertical $^1$MLCT energies from DMF to $H_2O$ covered in different solutions. Labels 1, 2 and 3 correspond to the complexes.

**Resonant Inelastic X-ray Scattering.** In order to establish the energetics of the MC excited states we utilize Fe 2p3d RIXS (Fig. 5). Different from UV-Visible absorption, the Raman selection rules for 2p3d RIXS results predominantly in scattering to MC excited states, providing direct access to the energies of these states. The RIXS spectra in Fig. 5A were recorded at the Fe $L_3$-edge X-ray absorption white-line resonance around 709 eV resulting from transitions from 2p to unoccupied 3d($e_g$) orbitals (see SI). The measured RIXS spectra of all three complexes are very similar. The spectra are dominated by an inelastic scattering feature at ~3 eV energy transfer, indicated by the light blue line in Fig 5A. Weak elastic scattering is clearly visible at 0 eV energy transfer. At higher energy transfer, there is a broad feature (centered at 6 – 7 eV) that can be assigned to decays from nominally ligand orbitals, thus the final states correspond to ligand-to-metal charge transfer (LMCT) excited states.[51,72] For the purposes of this work, we focus on the energies of the MC states observed in the RIXS spectrum and in particular to the dominant MC RIXS feature at 3 eV. To accurately obtain these energies we carried out the peak assignment and fitting described below.

Assignment of the ~3 eV MC RIXS feature is based on the Tanabe-Sugano energy matrices[73] and on previous RIXS studies of similar low spin $Fe^{2+}$ complexes.[51,72] Within the octahedral approximation, these allow the assignment of the dominant RIXS spectral peak unambiguously to the $^1T_{1g}(t_{2g}^5 e_g^1)$ excited state. Based on the Tanabe-Sugano energy matrices and the measured TES spectrometer instrument response, we fit the RIXS spectrum using only the octahedral ligand field 10Dq as a free fit parameter to describe the MC state energies (Racah parameters B and C are known, see SI for a detailed description of the analysis). The fit establishes that the energy transfer position of the dominant MC RIXS feature is a robust and accurate measure of $^1T_{1g}$ excited state energy (Fig. 5A). Additionally, this allows us to calculate the energies of other MC states, including the lowest energy $^3$MC states that are relevant for the electron back-transfer. We label each MC state using approximate octahedral notation and therefore the states discussed here are orbitally triply-degenerate. Within the more accurate $C_{2v}$ point group symmetry of the molecules, this degeneracy is lifted on the order of 0.1 eV.

Table 2. Summary of the vertical MC state energies derived from the fitting of the RIXS spectra of **1**, **2** and **3** (Fig. 5). The scaling factor for 3d electron repulsion Racah parameters is 75%: B = 0.110 eV and C = 0.406 eV (see SI for details).

| Complex | 10Dq (eV) | $^3T_{1g}$ (eV) | $^3T_{2g}$ (eV) | $^1T_{1g}$ (eV) | $^1T_{2g}$ (eV) |
|---|---|---|---|---|---|
| 1 | 3.24 | 2.09 | 2.80 | 2.98 | 4.40 |
| 2 | 3.22 | 2.08 | 2.78 | 2.97 | 4.38 |
| 3 | 3.19 | 2.04 | 2.75 | 2.93 | 4.34 |
| average | 3.22 | 2.07 | 2.78 | 2.96 | 4.38 |

The resulting MC excited state energies extracted from the fitting of the RIXS spectra are summarized in Table 2. The MC state energies of the different complexes are identical within the experimental uncertainties; therefore, all three ligands have the same ligand field. We take the averages of these as the best estimate for all three complexes. Thus, any changes in the MLCT lifetime between the complexes cannot be related to differences in Fe-ligand bonding, as it is effectively identical in all three complexes. The value of 10Dq, as expected, lies between [Fe(bpy)$_3$]$^{2+}$ (10Dq ≈ 2.8 eV[74]) and [Fe(CN)$_6$]$^{4-}$ (10Dq = 4.2 eV[75]). The lowest energy MC state is $^3T_{1g}$ at 2.07±0.1 eV, which is ~0.7 eV lower than the second lowest MC state $^3T_{2g}$. With these vertical MC states energies from Table 2, we can establish the relative energetics of the low energy MC excited states with respect to the MLCT states determined from the UV-Visible absorption (Fig. 5B). The shaded area in Fig. 5B shows the range of $^1$MLCT energies in various solvents due to the solvatochromic effect. Clearly, the $^1$MLCT energies are within the same range of the $^3T_{1g}$ energy, indicating a possibility of effective $^3$MLCT→$^3T_{1g}$ relaxation pathway.

## *Discussion*

Figure 6 shows the MLCT lifetime dependence on the $^1$MLCT minimum energy $E_0$. It is evident that complexes **1** and **2** show a very similar behaviour, expressed by an exponential increase in lifetime as $E_0$ is decreased. In contrast, complex **3** behaves qualitatively differently and shows a non-monotonic MLCT lifetime dependence on $E_0$. The appearance of a maximum in the lifetime clearly demonstrates that two independent MLCT relaxation pathways occur for **3**. The rate of one pathway becomes slower as the reactant MLCT energy decreases (normal Marcus region, similar to **1** and **2**), whereas the second pathway becomes faster (inverted Marcus region behaviour). We assign these two pathways to $^3$MLCT→$^3$MC (indirect) and $^3$MLCT→GS (direct), similar to the two relaxation mechanisms observed in Ru cyano-polypyridyl complexes by Indelli *et al.*[24] The total rate of intramolecular electron transfer is thus the sum of these two pathways:

$$k = k_{3MC} + k_{GS}.$$

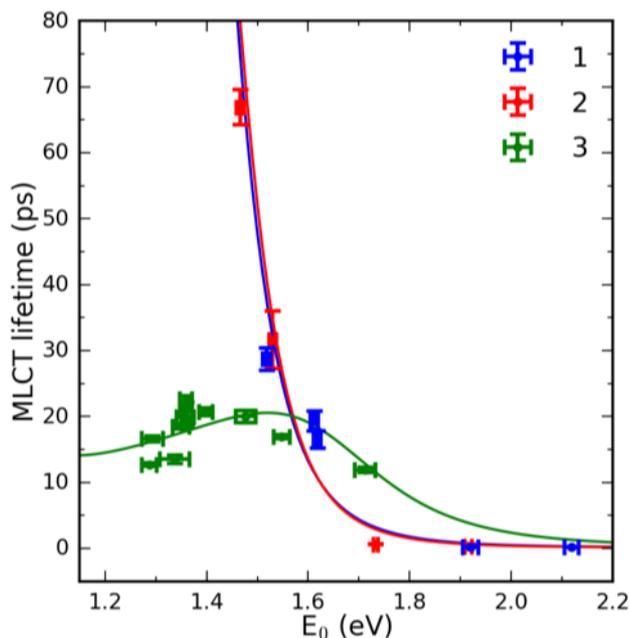

Fig. 6. Measured electron transfer lifetime dependence on the $^1$MLCT minimum energy $E_0$ of **1**, **2** and **3** (points). Solid lines correspond to calculated electron transfer lifetimes. See text for further details.

The indirect $^3$MLCT→$^3$MC pathway is clearly dominating the electron back-transfer in complexes **1** and **2**, based on the observation of $^3$MC intermediate in the TA and normal-Marcus behavior of the $^3$MLCT lifetimes. Thus, we first discuss in detail the indirect pathway in these two complexes. Following that, we include the direct pathway and focus on the dynamics of complex **3**.

**$^3$MLCT→$^3$MC electron transfer pathway.** In Fig. 7 we show a Marcus plot of the MLCT relaxation rates in **1** and **2**. Qualitatively, the observed rate constants follow a classical Marcus curve with pre-exponential factor $1/k_0$=150±40 fs and reorganization energy $\lambda_{cl,ET}$=0.43±0.1 eV (temperature is fixed at 298 K). We use the notation $\lambda_{cl,ET}$ to emphasize that the electron transfer reorganization energy in the Marcus model is classical. The driving force is calculated with $\Delta G = E_{shift} - E_0$, where $E_{shift}$ is a fit parameter with a value of 1.59±0.06 eV. The maximum position of the Marcus parabola (-$\Delta G$=$\lambda_{cl,ET}$, i.e. no barrier) at $E_0$=$E_{shift}$+$\lambda_{cl,ET}$=2.02±0.12 eV is within the range of the vertical $^3$T$_{1g}$ energy derived from the RIXS measurements. It is thus clear that MLCT depopulation is governed by a relaxation to the $^3$T$_{1g}$ manifold, as previously observed for **1** in H$_2$O.[35] Electron transfer to the $^3$MC state occurs from the $^3$MLCT state, as the $^1$MLCT→$^3$MLCT intersystem crossing in both Ru and Fe polypyridyl complexes occurs in less than 100 fs.[67,76] For instance, the $^1$MLCT→$^3$MLCT time constant was found to be <20 fs for [Fe(bpy)$_3$]$^{2+}$ in a fluorescence upconversion experiment.[67] As we argued in the Results section, the $^1$MLCT-$^3$MLCT energy gap $E_{ST}$ is about 0.25 eV.[70] Following these assignments, we get that $E_{shift}$=$E_0$($^3$MC)+$E_{ST}$, and therefore the minimum energy of the product $^3$MC state is $E_0$($^3$MC) ≈ 1.34 eV. In addition, $\lambda_{cl,ET}$=0.43±0.1 eV agrees reasonably well with the typical intramolecular reorganization energies of the $^3$MC states in Fe polypyridyl complexes (0.4–0.5 eV for [Fe(bpy)$_3$]$^{2+}$).[74] This reorganization energy is dominated by the displacements of low-frequency ($h\nu$<$k_BT$) modes associated with changes in metal-ligand bonding, including pseudo-Jahn-Teller tetragonal distortion and symmetric Fe-ligand bond expansion, thus justifying the application of the classical Marcus formula.

It is somewhat surprising that the MLCT lifetimes in **1** and **2** can be qualitatively described by a classical Marcus formula. In addition to the assumption that the nuclear dynamics is statistical with all the nuclear degrees of freedom in equilibrium during all observed time scales (certainly not valid for sub-picosecond MLCT lifetimes), the above analysis also assumes that electron transfer proceeds only from a cooled $^3$MLCT state with T=298 K. This is a reasonable assumption for MLCT lifetimes >10 ps, as the observed cooling lifetimes of the MLCT state is on the order of a few picoseconds. Wavepacket dephasing, intramode thermalization and intramolecular vibrational energy distribution (IVR) timescales are of the same order of magnitude or faster. We do not observe any coherent wavepacket dynamics even for electron transfer time scales faster than the Fe-ligand breathing frequency (**1** and **2** in H$_2$O and MeOH). This is different from the dynamics of higher-symmetry Fe polypyridyl complexes such as [Fe(bpy)$_3$]$^{2+}$.[77–79] Due to lower symmetry and stronger interaction with the solvent, the dephasing and cooling lifetimes in **1**, **2** and **3** complexes are likely faster. In summary, we expect that non-equilibrium effects are relevant for electron transfer lifetimes <10 ps. However, it seems that deviations from the equilibrium Marcus model in the barrierless region are not dramatic (factor of 2–3).

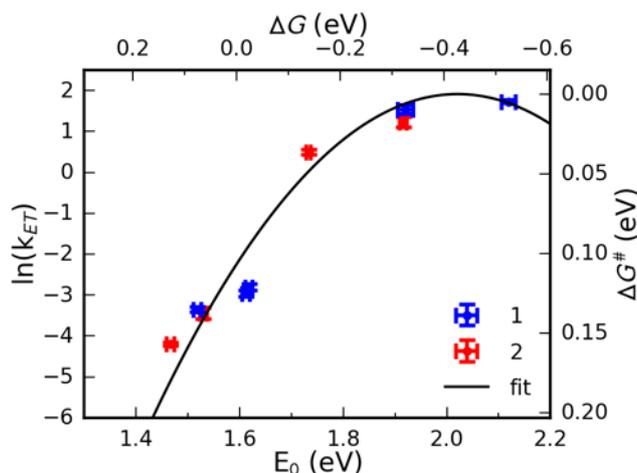

Fig. 7. Marcus plot of **1** and **2** MLCT rates. Black line corresponds to a fit of the classical Marcus formula with $1/k_0=150\pm40$ fs, $\lambda_{cl,ET}=0.43\pm0.1$ eV and T=298 K and. The energy shift relating $E_0$ to the driving force is $1.59\pm0.06$ eV (see text for further details). The top axis shows the driving force and right axis shows the energy barrier of the electron transfer.

In addition to the statistical nature of the nuclear dynamics, the Marcus analysis presented in Fig. 7 makes also several assumptions regarding the potential energy surfaces (PES) of the $^3$MLCT and $^3$MC states. Firstly, the analysis assumes that changes in $E_0$ correspond to changes in the driving force (i.e. $E_0(^3MC)$ and $E_{ST}$ do not change) and that $\lambda_{cl,ET}$ is constant in all solutions. The former assumption is likely approximately true, because both $E_0(^3MC)$ and $E_{ST}$ are dominated by the intramolecular interactions. Based on the discussion in Kjær et al. on the solvent dependence of the $^3$MC→GS lifetime in **1**,[35] and a comparison of the RIXS spectrum of **1** measured from a solid sample and from a H$_2$O solution (see SI), we find that the changes in the $^3$MC energy are most likely around 0.1 eV. Such changes have only a small effect to the observed MLCT lifetimes in H$_2$O because the molecule remains in an effectively barrierless region. In contrast, the assumption that $\lambda_{cl,ET}$ is constant is likely not strictly valid,particularly because analysis of the UV-Visible absorption spectra revealed a considerable solvent dependence of the MLCT reorganization energies. However, there are several different contributions to $\lambda_{cl,ET}$ of the indirect $^3$MLCT→$^3$MC pathway. The largest contribution is likely due to intramolecular Fe-ligand bond expansions, and that would explain the relative $\lambda_{ET}$ insensitivity from solvent. Secondly, we note that Marcus analysis in general assumes that only displacement of the harmonic PESs determines the barrier height and any higher order vibronic effects (change in vibrational frequencies, Duschinsky rotation, anharmonicities) are negligible. It is known from computational studies of the Fe polypyridyl spin crossover complexes that the frequencies of Fe-ligand stretching vibrations are softer in $^{3,5}$MC states in comparison to the GS and $^{1,3}$MLCT states. In [Fe(bpy)$_3$]$^{2+}$, an 18 meV (145 cm$^{-1}$) stretching mode in the GS has been calculated to be 14 meV (116 cm$^{-1}$) in the $^5$MC state.[80] It is also unlikely that the $^3$MLCT and $^3$MC PESs are perfectly harmonic in the whole nuclear space sampled by the molecule, given the large driving forces and reorganization energies in the order of ~0.5 eV. However, it has been argued for the $^5$MC->GS transition in Fe spin crossover complexes that these higher order vibronic effects have a tendency to cancel out (see Scheme 3 by Hauser[81]). It is therefore likely that the magnitude of these effects is smaller than the errors due to inaccuracies in determination of the driving forces and reorganization energies.

**$^3$MLCT→GS electron transfer pathway.** Evaluation of the $^3$MLCT→GS rates is more straightforward than the $^3$MLCT→$^3$MC rates, because the relevant driving force and reorganization energies can be extracted from the UV-Visible absorption spectrum within the Condon approximation. The $^3$MLCT→GS pathway is in the inverted Marcus regime and therefore quantum mechanical effects of the high-frequency vibrational modes need to be considered. To include the latter effects, we utilize a rate formula with a single quantum mode and a classical mode:[40]

$$k = k_0 \sum_n \frac{S^n e^{-S}}{n!} exp\left(-\frac{(\Delta G + nh\upsilon + \lambda_{cl,ET})^2}{4\lambda_{cl,ET} k_B T}\right).$$

The Huang-Rhys parameter is $S=\lambda_q/h\upsilon$. In case of the $^3$MLCT→GS pathway, $\Delta G=-E_0+E_{ST}$, $\lambda_{cl,ET}=\lambda_{cl}$ (i.e. classical reorganization energy of the direct pathway is equal to the classical MLCT reorganization energy retrieved from the UV-Visible absorption spectra). We use the same formula to also calculate the $^3$MLCT→$^3$MC rates and in that case $\Delta G= E_0(^3MC)-E_0+E_{ST}$ and $\lambda_{cl,ET}=\lambda(^3MC)+\lambda_{cl}$. For both pathways, changes in $\lambda_q$ and $\lambda_{cl}$ in different solvents are incorporated via the linear fits in Fig. 4. $\lambda(^3MC)$ is a constant that describes the reorganization energy of the $^3$MC state due to Fe-ligand expansion.

Results of this modelling with two pathways are shown in Fig. 6 for the three complexes. Four free parameters were used to adjust the calculated MLCT lifetimes to the experiment, and all of these are same for **1**, **2** and **3**. The adjusted parameters are the two pre-exponential factors $k_{3MC,0}$=1/100 fs$^{-1}$ and $k_{GS,0}$=1/120 fs$^{-1}$ and $E_0(^3MC)$ = 1.2 eV and $\lambda(^3MC)$ = 0.5 eV. With these parameters, we can describe the MLCT lifetime trends in all the investigated solutions of the three complexes. The model used here is similar to the approach developed by Barbara *et al.* to simulate femto- and picosecond electron back-transfer in various solvated molecules.[66,82,83] However, our model is simpler as we found that inclusion of dynamical solvation effects is not necessary to describe the observed $^3$MLCT lifetimes.[84] Note that we have also omitted a weak $1/\sqrt{\lambda}$ dependence of the pre-exponential factors and consider these as constants.

Given the several assumptions in the modelling and uncertainties in the parameters, the calculated lifetime curves in Fig. 6 are likely not unique. However, we can reliably draw some qualitative conclusions. First, inclusion of the $^3$MLCT→GS pathway explains the difference between complex **3** and complexes **1** and **2**. The $^3$MLCT→GS pathway is more prominent in **3** because of the larger $\lambda_q$ values in weakly interacting solvents (Fig. 4B). The slow change of the MLCT lifetimes at $E_0$<1.5 eV in **3** is due to compensation of the driving force decrease by the increase in $\lambda_{cl}$. The model predicts that complex **3** at $E_0$>1.6 eV has longer $^3$MLCT lifetimes than **1** and **2**. This is supported by the longer experimental MLCT lifetime of **3** in H$_2$O compared to **2** in MeOH at $E_0$=1.7 eV. Longer lifetime of **3** at $E_0$>1.6 eV be therefore explained by the larger $\lambda_{cl}$ value that increases the barrier for the $^3$MLCT→$^3$MC pathway.

The pre-exponential factor $k_{GS,0}$ = 1/120 fs$^{-1}$ = 8·10$^{12}$ s$^{-1}$ is about 10–50 times smaller than determined by Indelli *et al.*[24] for Ru dicyano-bipyridine and tricyano-terpyridine complexes. Given that $k_0 \propto |V|^2$, where V is electronic coupling between the reactant and product states, then this is roughly consistent with the smaller spin-orbit coupling in Fe complexes. 3d spin-orbit coupling of Fe is $\xi_{3d}$≈0.05 eV, three times smaller than the 4d spin-orbit coupling of Ru $\xi_{4d}$≈0.15 eV.[70] In contrast, $k_{3MC,0}$ = 1/100 fs$^{-1}$ = 10$^{13}$ s$^{-1}$ is three orders of magnitude larger in the Ru tricyano-terpyridine complexes. This indicates significantly smaller $^3$MLCT-$^3$MC electronic coupling in Ru complexes. In addition to ~1 eV higher $^3$MC energy, this seems to be another important factor why the indirect pathway is "blocked" in Ru complexes.

The presence of an effective $^3$MLCT→GS pathway at smaller $E_0$ energies limits the extent one can prolong the MLCT lifetimes by lowering the acceptor energy for a given ligand. Extrapolation of the calculated MLCT lifetime of **1** and **2** gives a maximum lifetime around ~200 ps at $E_0$ = 1.3–1.4 eV, followed by a rapid decrease in the lifetime at lower $E_0$ values. We discuss the possible strategies to extend the MLCT lifetime further in the Conclusion.

## *Conclusion*

We have explored the MLCT lifetime dependence of Fe cyano-polypyridyl complexes **1**, **2** and **3** in various solvents over a wide range of MLCT energies. We found that MLCT relaxation can proceed via two different electron transfer pathways: a direct $^3$MLCT→GS pathway and an indirect $^3$MLCT→$^3$MC pathway. The latter electron transfer takes place in a normal Marcus regime and can be qualitatively described with a classical Marcus formula. The reorganization energy for this pathway is large due to considerable intramolecular reorganization associated with the $^3$MC state. The direct $^3$MLCT→GS electron transfer takes place in an inverted Marcus regime, with the solvent and high frequency ($h\nu$ = 0.19 eV = 1530 cm$^{-1}$) polypyridyl skeleton mode being coupled to this transition. Dynamics of complex **3** in weakly interacting solvent therefore resembles Ru and Os polypyridyl photosensitizers where the direct pathway similarly dominates the $^3$MLCT relaxation.

Additionally, we have used RIXS measurements to directly probe the low-energy MC excited states that cannot be accessed with other conventional spectroscopic methods. These measurements corroborate the conclusion from prior measurements on **1** that the $^3$MC excited state product corresponds to the $^3$T$_{1g}$ manifold (octahedral notation). We used a novel TES array spectrometer, that although having a sub-optimal emission energy resolution, has >100 times higher detection efficiency than conventional soft X-ray grating spectrometers and can thus be utilized for experiments on dilute and radiation sensitive samples. This demonstrated the potential of RIXS for gaining insight into the photophysics and -chemistry of MC excited states of 3d transition metal complexes.

The modelling applied here identifies the key electronic structure parameters that govern the of MLCT relaxation rates. This model is of general importance in rationalizing the MLCT lifetimes in Fe coordination complexes. The longest MLCT lifetime in a ferrous complex was reported so far by Chabera *et al*.[20] The complex there has a low energy $^1$MLCT state around ~800 nm, and the lifetime of 528 ps is in the range that could be observed in the complexes investigated here at similar MLCT energies without the direct pathway. These considerations could be contrasted to the record 2 ns $^2$LMCT lifetimes in a ferric complex observed by Kjaer *et al*.[21] The reported Arrhenius analysis revealed an almost order-of-magnitude smaller energy barrier ($\Delta G^{\neq}$ = 0.03 eV) than observed here for the indirect pathways in the solution with longest $^3$MLCT lifetime ($\Delta G^{\neq}$~0.15 eV, **2** in DMF). This is consistent with the relatively high energy of the LMCT state (~500 nm). Remarkably, the pre-exponential factor in the complex studied by Kjær *et al*. is 1 ns, as opposed to ~0.1 ps observed in the ferrous complexes investigated here. This suggests that the key factor behind the record lifetimes in ferric carbene complexes is the spin-forbidden two-electron $^2$LMCT-$^4$MC electronic coupling, which is apparently 100 times smaller than the spin-allowed one-electron $^3$MLCT-$^3$MC electronic coupling in similar ferrous complexes.

We can outline specific guidelines for increasing the MLCT lifetimes in ferrous complexes based on our investigation. At low $^1$MLCT energies ($E_0$<1.5 eV), suppression of the direct pathway by decreasing the solvent and ligand related MLCT reorganization energies is important for achieving long MLCT lifetimes. Here, synthetic strategies previously devised for 4d and 5d polypyridyl complexes could be applied. These strategies generally aim to decrease structural displacements of the polypyridyl ligand upon MLCT excitation, thus decreasing the Huang-Rhys parameter of the respective high-frequency vibrational modes. This is accomplished either by increasing delocalization of the excited electron over a larger ligand framework, or by increasing rigidity of the ligand with chemical links between pyridyl rings, as reviewed for a wide range of ligands of Ru and Os complexes by Treadway et al.[85] At higher $^1$MLCT energies ($E_0$>1.5 eV), where the indirect $^3$MLCT→$^3$MC pathway dominates, increasing the MLCT and $^3$MC reorganization energy would extend the MLCT lifetime. Increasing the ligand field further (increasing the energy of $^3$MC) would also suppress the indirect pathway.

*Acknowledgements*

This work was supported by the U.S. Department of Energy, Office of Science, Basic Energy Sciences, Chemical Sciences, Geosciences, and Biosciences Division. Use of the Stanford Synchrotron Radiation Lightsource, SLAC National Accelerator Laboratory, is supported by the U.S. Department of Energy, Office of Science, Office of Basic Energy Sciences under Contract No. DE-AC02-76SF00515. We gratefully acknowledge Joe Fowler, Gene Hilton, Carl Reintsema, and Dan Schmidt in the NIST Quantum Sensors Project for their contribution to the development of the TES spectrometer.